\documentstyle[prl,twocolumn,aps]{revtex}

\begin{document}

\draft


\twocolumn
[\hsize\textwidth\columnwidth\hsize\csname@twocolumnfalse\endcsname
\title{Localized $4f$ States and Dynamic Jahn-Teller Effect in PrO$_2$}

\author{A.T. Boothroyd$^1$, C.H. Gardiner$^1$, S.J.S. Lister$^1$, P. Santini$^1$, B.D. Rainford$^2$, L.D. Noailles$^3$, D.B. Currie$^4$,\\ R.S. Eccleston$^5$, R.I. Bewley$^5$ }

\address{$^1$ Department of Physics, University of Oxford, Oxford, OX1 3PU,
United Kingdom\\ $^2$ Department of Physics, University of
Southampton, Southampton, SO17 1BJ, United Kingdom\\ $^3$
Inorganic Chemistry Laboratory, University of Oxford, Oxford, OX1
3QR, United Kingdom\\ $^4$ Department of Chemistry, University of
Southampton, Southampton, SO17 1BJ, United Kingdom \\ $^5$ ISIS
Facility, Rutherford Appleton Laboratory, Chilton, Didcot, OX11
0QX, United Kingdom}

\date{\today}
\maketitle

\begin{abstract}
Neutron spectroscopic measurements of the magnetic excitations in
PrO$_2$ reveal (1) sharp peaks characteristic of transitions
between levels of the $4f^1$ configuration of Pr$^{4+}$ split by
the cubic crystal field, and (2) broad bands of scattering centred
near 30\,meV and 160\,meV. We present a simple model based on a
vibronic Hamiltonian that accounts for the main features of the
data. The analysis shows that $90\pm10$\% of the Pr ions have a
localized $4f^1$ configuration and provides strong evidence for a
dynamic Jahn-Teller effect in the $\Gamma_8$ electronic ground
state.
\end{abstract}

 ]

Anomalous static and dynamic magnetic properties of rare earth and
actinide compounds containing strongly correlated electrons are
usually attributed to Kondo or valence fluctuations, or to the
formation of heavy fermion states. If there are unquenched orbital
degrees of freedom, however, then other mechanisms for anomalous
behavior are possible. One such mechanism is the coupling of
electronic charge fluctuations to phonons, and leads to
magnetoelastic and Jahn-Teller phenomena. In the case of localized
electrons the dynamic Jahn-Teller effect (DJTE) may quench
electronic degrees of freedom, causing a reduction in the magnetic
moment and forming states of mixed magnetic and phonon character
\cite{Bersuker}.

Electron-lattice effects of this kind are of course well known but
the absence of a clean reference system has undoubtedly hindered
their identification among other anomalous properties, especially
in Ce and U compounds. In this Letter we show that PrO$_2$
represents such a model DJTE system. We find that magnetoelastic
coupling has a dramatic influence on the static and dynamic
properties of PrO$_2$, and we provide the first direct
measurements of the magnetovibrational spectrum in a rare earth
DJTE system. We also show that the Pr ions exist almost entirely
in a localized $4f^1$ configuration, in disagreement with an
intermediate valence model for PrO$_2$.

PrO$_2$ is an insulator with the fluorite structure type. It
exhibits type-I antiferromagnetic ordering below $T_{\rm N} =
14$\,K with an anomalously low ordered moment of
$\sim$0.6\,$\mu_{\rm B}$ \cite{Kern-SSC-1984}. The electronic
ground state of PrO$_2$ has,
together with 
CeO$_2$ and TbO$_2$, been controversial for many years
\cite{Kotani-AP-1988,Bianconi-PRB-1988,Wuilloud-PRL-1984,Karnatak-PRB-1987}.
Core-level photoabsorption and photoemission spectra have been
interpreted by some authors
\cite{Kotani-AP-1988,Bianconi-PRB-1988} in terms of an
intermediate valence ground state comprised of a roughly 50:50
admixture of localized (i.e. atomic-like) $4f^1$ and $4f^2$
states. Others \cite{Wuilloud-PRL-1984,Karnatak-PRB-1987},
however, believe in a localized $4f^1$ configuration with a degree
of covalent mixing such that the oxygen $2p$ valence band contains
some extended states of $f$ symmetry. Publication of an
influential model \cite{FR-PRL-1993} describing an intermediate
valence state in the anomalous cuprate PrBa$_2$Cu$_3$O$_{6+x}$
(PrBCO) similar to that proposed for the lanthanide dioxides has
increased interest in PrO$_2$, making it a key reference compound
in high energy spectroscopic studies of PrBCO
\cite{Neukirch-Lytle-Hu-Staub} and related materials.

The aim of the present work was to clarify the electronic ground
state of PrO$_2$. Neutron spectroscopy probes electronic
transitions within the ground state configuration both within and
between the $^{2S+1}L_{J}$ terms, including splittings due to the
crystalline electric field (CEF) or exchange fields. The low
neutron energies ($\lesssim1$\,eV) and weak coupling to the
electronic angular momentum ensure that final state effects are
insignificant. The experiments were performed on the High Energy
Transfer (HET) chopper spectrometer at the ISIS Facility. On HET
neutrons of well-defined incident energy $E_{\rm i}$ are delivered
to the sample in short pulses. Spectra are recorded as a function
of flight time in banks of detectors surrounding the incident beam
direction. Because the spectra are recorded at constant scattering
angle $\phi$ the scattering vector $Q$ varies with energy $E$
thus: $\hbar^2Q^2/2m = 2E_{\rm i} - E - 2\cos\phi\sqrt{E_{\rm
i}(E_{\rm i}-E)}$.

Polycrystalline samples of PrO$_2$ were prepared by oxidation of
commercially obtained Pr$_6$O$_{11}$. The starting material was
baked in air at 1000\,$^{\circ}$C for several hours and then
annealed either in flowing O$_2$ at 280\,$^{\circ}$C for
approximately 20 days or under 300--500\,atm pressure of O$_2$ at
350\,$^{\circ}$C for 3--5 days. The products were checked by x-ray
diffraction and no trace of residual Pr$_6$O$_{11}$ could be
detected. From this, and a comparison of our neutron data with
that of Pr$_6$O$_{11}$ \cite{HM-ZPB-1992}, we estimate an upper
limit of 1\% on the amount of Pr$_6$O$_{11}$ impurity in our
samples.

For the neutron measurements approximately 10\,g of PrO$_2$ was
wrapped in aluminium foil and mounted in contact with the cold
head of a closed-cyle refrigerator. Data were collected with
$E_{\rm i} = 30$, 180, 450, 750 and 1200\,meV, and the scattering
from vanadium was used to normalize the spectra and to convert
them into units of absolute scattering cross-section. The
presented spectra correspond to the partial differential
cross-section ${\rm d}^2\sigma/{\rm d}\Omega{\rm d}E_{\rm f}$
multiplied by the factor $k_{\rm i}/k_{\rm f}$ \cite{Lovesey},
where $k_{\rm i}$ and $k_{\rm f}$ are the initial and final
neutron wavevectors and $E_{\rm f}$ is the final energy.
Measurements were made under identical conditions on a 10\,g
sample of polycrystalline CeO$_2$ in order to estimate the
non-magnetic (phonon + multiple scattering) background in the
PrO$_2$ data. CeO$_2$ is an almost ideal material for this purpose
as it has the same structure as PrO$_2$ and the scattering
cross-sections of Pr and Ce differ by only 10\%. The estimated
transmission of the PrO$_2$ and CeO$_2$ samples was 93--97\%
(depending on $E$ and $E_{\rm i}$), and the spectra were corrected
accordingly.

Fig.\ 1(a) shows spectra of PrO$_2$ and CeO$_2$ with $E_{\rm
i}=180$\,meV averaged over the angular ranges 9--19$^{\circ}$
($\langle \phi \rangle = 14^{\circ}$) and 125--139$^{\circ}$
($\langle \phi \rangle = 133^{\circ}$). At high angles the
scattering is dominated by one-phonon processes, and so the
$133^{\circ}$ data measures the phonon density of states weighted
by the neutron cross-section of the atoms. The $133^{\circ}$ data
for the two compounds are similar except that some PrO$_2$ peaks
are shifted to lower energies relative to CeO$_2$, most noticeably
below 30\,meV. At low angles more marked differences are observed.
The PrO$_2$ data contains a sharp peak at 131\,meV and a broad
peak centred on $\sim$30\,meV extending from 10\,meV to 80\,meV,
whereas the CeO$_2$ spectrum shows little structure and lies below
the PrO$_2$ spectrum at all energies. The 131\,meV peak has been
observed previously \cite{Kern-SSC-1984}. Both the broad and sharp
peaks in the PrO$_2$ data decrease in intensity with increasing
$Q$ consistent with a typical $4f$ magnetic form factor, and so we
identify these features as magnetic in origin. Conversely, the
CeO$_2$ signal systematically increases in intensity with $Q$, and
this implies that the CeO$_2$ scattering is non-magnetic as
expected for a $4f^0$ configuration with no admixture of localized
$4f^1$ states.

Several other magnetic features were observed in the PrO$_2$ data
in addition to those already described. The runs with $E_{\rm
i}=750$\,meV and 1200\,meV detected two more peaks. These are
shown in Figs.\ 1(b) and (c). The peak centred near 350\,meV in
the 750\,meV run is significantly broader than the resolution,
whereas the peak centred at 730\,meV in the 1200\,meV run is
resolution-limited. The 450\,meV run revealed a shoulder of
scattering above the 131\,meV peak, centred at $\sim$160\,meV.
This feature is displayed in the inset to Fig.\ 1(d). Finally, the
$E_{\rm i}=30$\,meV run revealed a peak centred at 3\,meV shown in
Fig.\ 1(d). This peak shifted to lower energies as the temperature
was raised, becoming quasielastic above $\sim$15\,K. Given that
$T_{\rm N} = 14$\,K we can confidently attribute this 3\,meV peak
to spin wave excitations of the antiferromagnetically ordered
ground state. We found no evidence in our data for an excitation
at 5.4\,meV reported in a previous work \cite{Kern-JAP-1990}, and
so we conclude that this feature is not intrinsic to PrO$_2$.

Fig.\ 1(d) shows data from several runs after correction for the
non-magnetic scattering and for the calculated (see below)
magnetic form factor. The form factor correction extrapolates the
data to zero $Q$ and hence allows comparison of runs with
different incident energies.

We begin our interpretation of the results by considering the
effect of the CEF. The splitting of a localized $4f^1$
configuration in a cubic CEF is described by the spin-orbit
coupling constant $\zeta$, which sets the
$^2F_{5/2}\rightarrow{^2F_{7/2}}$ separation, and the fourth and
sixth order CEF parameters $B_{0}^{4}$ and $B_{0}^{6}$. Using the
free ion value of $\zeta$ for Pr$^{4+}$ \cite{Kaufman-JRNBS-1967}
Kern {\it et al} \cite{Kern-PRB-1985} have calculated the energy
levels as a function of the ratio $B_{0}^{6}/B_{0}^{4}$. They
assigned the 131\,meV peak to the $\Gamma_{8}\rightarrow\Gamma_7$
transition of $^2F_{5/2}$, and by estimating the ratio
$B_{0}^{6}/B_{0}^{4}$ from a point charge model they arrived at a
parameterization of the CEF. Kern {\it et al}'s model predicts the
levels in the CEF-split $^2F_{7/2}$ term to be $\sim$320\,meV
($\Gamma_{6}'$), $\sim$390\,meV ($\Gamma_{8}'$) and $\sim$580\,meV
($\Gamma_{7}'$) above the ground state in rough accord with the
peaks in Figs.\ 1(b) and (c). We conclude, therefore, that these
high energy peaks arise from the $^2F_{5/2}\rightarrow{^2F_{7/2}}$
intermultiplet transitions, the only proviso being that the signal
in Fig.\ 1(b) must encompass both the $\Gamma_{6}'$ and
$\Gamma_{8}'$ levels. Assuming this to be the case we refined the
CEF model with the 14 states $|J;m_{J}\rangle$ of $^2F_{5/2}$ and
$^2F_{7/2}$ as a basis. With all 3 parameters allowed to vary
independently the fitting procedure converged yielding
$B_{0}^{4}=(-776\pm8)$\,meV, $B_{0}^{6}=(207\pm5)$\,meV and $\zeta
= (100.5\pm1)$\,meV \cite{footnote2} and gave excellent agreement
with the observed energies.

With the eigenfunctions from the CEF refinement we can now
evaluate the cross-sections for the observed transitions as a
function of $Q$. The formulae \cite{Lovesey} contain radial
moments of the $4f$ wavefunction, and values for Pr$^{4+}$ were
taken from Ref. \cite{PJB-WBL}. The curve in Fig.\ 1(b) represents
the calculated cross-sections for the
$\Gamma_8\rightarrow\Gamma_{6}'$ and
$\Gamma_8\rightarrow\Gamma_{8}'$ transitions, and the curve on
Fig.\ 1(c) is likewise for the $\Gamma_8\rightarrow\Gamma_{7}'$
transition. Considering that no additional parameters are needed
to obtain the calculated cross-sections the level of agreement
with the data is very good. The same cannot be said, however, when
the cross-sections for the $\Gamma_8\rightarrow\Gamma_8$ and
$\Gamma_8\rightarrow\Gamma_7$ transitions are calculated. At zero
$Q$ these are 182 and 99\,mb\,sr$^{-1}$Pr$^{-1}$ respectively
\cite{footnote3}, considerably greater than the integrated
spectral weights 47 and 48\,mb\,sr$^{-1}$Pr$^{-1}$ in the 3\,meV
and 131\,meV peaks shown in Fig.\ 1(d) \cite{footnote4}.

We will now describe a model that both accounts for the puzzling
intensities just mentioned and explains the origin of the
10--80\,meV broad peak and the 160\,meV shoulder. Our hypothesis
is that these effects derive from a strong coupling between $4f^1$
electronic states and local dynamic lattice distortions. This
coupling mixes electronic and phonon degrees of freedom and causes
a dynamic Jahn-Teller effect in the $\Gamma_{8}$ ground state.

The existence of a DJTE in PrO$_2$ was in fact suggested some time
ago to explain the factor $\sim$2 reduction in ordered moment
relative to a pure $\Gamma_8$ ground state \cite{Kern-SSC-1984}.
An analogous moment reduction in UO$_2$ had already been
attributed to a DJTE \cite{Sasaki}, and so its occurrence in
PrO$_2$ would not be a surprise. Moreover, the neutron data in
Fig.\ 1(a) showing substantial phonon shifts in PrO$_2$ relative
to CeO$_2$ provide direct evidence for a significant
magnetoelastic interaction. To find out how this interaction
influences the magnetic excitation spectrum we describe a simple
model for the coupling of a set of CEF levels with phonons. Group
theory shows that the $\Gamma_8$ CEF ground state couples in first
order to local distortions of either $\Gamma_3$ or $\Gamma_5$
symmetry, and point charge calculations indicate that the coupling
strengths are comparable \cite{Sasaki}. For the sake of
simplicity, however, we will restrict our model to three
degenerate local distortions of $\Gamma_5$ symmetry with
vibrational frequency $\omega_{\rm ph}$. We describe the
magnetoelastic coupling with the Hamiltonian
\begin{equation}
H_{\rm ME} = g\sum_{i} (a_{i} + a_{i}^{\dag}) O_{i},
\label{eq:HME}
\end{equation}
where $g$ is the coupling constant, $a_{i}$ and $a_{i}^{\dag}$ are
phonon annihilation and creation operators, and $O_{i}$ are the
$\Gamma_5$ symmetry quadrupolar operators: $O_1 =
(J_yJ_z+J_zJ_y)/2$, $O_2 = (J_zJ_x+J_xJ_z)/2$, and $O_3 =
(J_xJ_y+J_yJ_x)/2$. To investigate the ordered magnetic moment we
also included a molecular exchange field interaction $H_{\rm
Ex}={\bf H}\cdot{\bf J}$. The magnitude of $\bf H$ was chosen so
as to reproduce the observed 3\,meV splitting of the ground state.

We take as a basis the 24 states $|\phi_n\rangle$,
$n=1,\ldots,24$, represented by $|\Gamma_8;0\rangle$,
$|\Gamma_7;0\rangle$, $|\Gamma_8;1^{(i)}\rangle$, and
$|\Gamma_7;1^{(i)}\rangle$. These are products of the CEF
eigenstates and the three $\Gamma_5$ phonon modes ($i=1$, 2, 3)
containing either 0 or 1 quanta $\hbar\omega_{\rm ph}$. The energy
separation $\Delta$ of $|\Gamma_8;0\rangle$ and
$|\Gamma_7;0\rangle$ is chosen so as to reproduce the observed
peak at 131\,meV. The vibronic states are obtained by
diagonalisation of the matrix $\langle \phi_{n'}|H_{\rm ME}+H_{\rm
Ex}|\phi_n \rangle$, and the magnetic part of the neutron
cross-section is calculated from the matrix elements of ${\bf J}$.

We evaluated the model for a range of parameters. Small values of
$\hbar\omega_{\rm ph}$ are effective at mixing
$|\Gamma_8;0\rangle$ and $|\Gamma_8;1^{(i)}\rangle$ states
creating a DJTE. The consequences are a transfer of intensity from
the ground state into the phonon-like states and a reduction in
the ordered moment. A large $\hbar\omega_{\rm ph}$ tends to create
vibronic states involving the $\Gamma_7$ CEF excitation. To
illustrate these effects we plot on Fig.\ 2 the zero-$Q$ energy
spectrum for two parameter sets, one (I) with small and the other
(II) large $\hbar\omega_{\rm ph}$. In both cases new peaks appear
between the 3\,meV and 131\,meV levels. With set I there is a
factor 0.86 reduction in the ordered moment (independent of the
direction of $\bf H$) and a similar decrease in the intensity of
the 3\,meV peak. Set II produces a smaller effect on the ordered
moment and 3\,meV peak, but a new feature is the signal above the
131\,meV peak.

These results give us confidence that magnetoelastic coupling is
responsible for the unusual energy spectrum and reduced moment of
PrO$_2$. Because of its simplicity we cannot expect a
single-frequency model to match the data perfectly. In reality
there will be coupling to local dynamic distortions with
$\Gamma_3$ as well as $\Gamma_5$ symmetry, and these distortions
will exist over a range of frequencies due to dispersion. The
extent of the observed broad scattering is indicative that many
frequencies are actually involved. A realistic model would also
need to include states with $>$1 phonons. Nevertheless, the
qualitative agreement achieved with the present model is
satisfying.

Before concluding we will address the nature of the $f$ states.
Our analysis has assumed a localized $4f^1$ configuration. We can
check this assumption two ways. Firstly, we can look for the
signature of localized $4f^2$ states in the range 200--300\,meV
where the $^3H_4 \rightarrow {^3H_5}$ transition occurs
\cite{Sugar-PRL-1965}. No peaks are observed. Second, we can
directly determine the number of $4f^1$ states per Pr from the
neutron cross-section via the sum rule for transitions within a
$J$ multiplet \cite{Lovesey}. The measured zero-$Q$ cross-section
integrated up to 220\,meV, including the elastic scattering
\cite{footnote4}, is 240\,mb\,sr$^{-1}$Pr$^{-1}$. This compares
with the calculated value of $182 + 99 =
281$\,mb\,sr$^{-1}$Pr$^{-1}$ for the $J=\frac{5}{2}$ multiplet
\cite{footnote3}. Thus, by including the broad scattering in the
integral we recover almost all the missing intensity. Allowing a
10\% uncertainty in the absolute calibration we conclude that
$90\pm10$\% of the Pr ions have a localized $4f^1$ configuration.
The intermultiplet transitions shown in Figs.\ 1(b) and (c)
provide further support for this conclusion since the calculated
cross-sections for $4f^1$ are in good agreement with the measured
intensities.

In summary, our findings support a tetravalent model for PrO$_2$
with an almost fully occupied atomic $4f$ orbital. The neutron
data allow the possibility of $\sim$10\% under-occupancy relative
to a pure $4f^1$ configuration, and this may indicate a degree of
covalency. Any other occupied states of $f$ symmetry must then be
extended states in the O $2p$ valence band. These results rule out
the strongly intermediate valence model for PrO$_2$, and we
anticipate that a similar picture will apply to CeO$_2$ and
TbO$_2$. We find compelling evidence that magnetoelastic coupling
in PrO$_2$ creates a ground state with mixed electronic and
vibrational degrees of freedom by means of a DJTE. Being an
insulator with a simple structure, localized $f$ electrons and a
large CEF splitting, PrO$_2$ would seem an ideal model system for
further studies of the DJTE.

We thank Mike Hayward for help with sample preparation, and the
Engineering and Physical Sciences Research Council of Great
Britain for financial support. P.S. thanks the Swiss NSF for
financial support.


\begin{figure}
\caption{Neutron inelastic scattering from PrO$_2$ and CeO$_2$
measured at a temperature of 10\,K. In (a) the incident neutron
energy $E_{\rm i}$ was 180\,meV, and spectra recorded at low and
high scattering angles are shown. (b) and (c) show
$^2F_{5/2}\rightarrow{^2F_{7/2}}$ intermultiplet transitions
measured with $E_{\rm i} = 750$\,meV and 1200\,meV respectively.
The solid lines are cross-sections calculated from the CEF model
described in the text with widths corresponding to the
instrumental resolution. (d) shows data corrected for the
non-magnetic background and the $Q$ dependence of the magnetic
cross-section as described in the text. Main frame: $E_{\rm i} =
30$ and 180\,meV. Inset: $E_{\rm i} = 450$\,meV showing the
shoulder to the 131\,meV peak. }
\end{figure}

\begin{figure}
\caption{Zero-$Q$ cross-section of PrO$_2$ calculated from the
magnetoelastic model described in the text with parameter sets I
($\hbar\omega_{\rm ph}=12.5$\,meV, $g=8$\,meV and
$\Delta=109$\,meV) and II ($\hbar\omega_{\rm ph}=42$\,meV,
$g=9$\,meV and $\Delta=101$\,meV). In both cases ${\bf H}$ was
0.5\,meV $\parallel (1,1,1)$. The peak widths correspond roughly
with the experimental resolution in Fig.\ 1(d).}
\end{figure}


\begin{references}

\bibitem{Bersuker}
I.B. Bersuker and V.Z. Polinger, {\it Vibronic Interactions in
Molecules and Crystals}, (Springer-Verlag, 1989).

\bibitem{Kern-SSC-1984}
S. Kern {\it et al}, Solid State Commun. {\bf 49}, 295 (1984). The
direction of the ordered moment is still not known.

\bibitem{Kotani-AP-1988}
A. Kotani {\it et al}, Adv. Phys. {\bf 37}, 37 (1988).

\bibitem{Bianconi-PRB-1988}
A. Bianconi {\it et al}, Phys. Rev. B {\bf 38}, 3433 (1988); H.
Ogasawara {\it et al}, Phys. Rev. B {\bf 43}, 854 (1991); S.
Kimura {\it et al}, J. Electron Spectrosc. {\bf 78}, 135 (1996);
S.M. Butorin {\it et al}, J. Phys.: Condens. Matter {\bf 9}, 8155
(1997).

\bibitem{Wuilloud-PRL-1984}
E. Wuilloud {\it et al}, Phys. Rev. Lett. {\bf 53}, 202 (1984); F.
Marabelli and P. Wachter, Phys. Rev. B {\bf 36}, 1238 (1987).

\bibitem{Karnatak-PRB-1987}
R.C. Karnatak, {\it et al}, Phys. Rev. B {\bf 36}, 1745 (1987); H.
Dexpert {\it et al}, Phys. Rev. B {\bf 36}, 1750 (1987).


\bibitem{FR-PRL-1993}
R. Fehrenbacher and T.M. Rice, Phys. Rev. Lett. {\bf 70}, 3471
(1993).

\bibitem{Neukirch-Lytle-Hu-Staub}
U. Neukirch {\it et al}, Europhys. Lett. {\bf 5}, 567 (1988); F.W.
Lytle {\it et al}, Phys. Rev. B {\bf 41}, 8955 (1990); Z. Hu {\it
et al}, Phys. Rev. B {\bf 60}, 1460 (1999); U. Staub {\it et al},
Phys. Rev. B {\bf 61}, 1548 (2000).



\bibitem{HM-ZPB-1992}
E. Holland-Moritz, Z. Phys. B {\bf 89}, 285 (1992).

\bibitem{Lovesey}
S.W. Lovesey, {\it Theory of Neutron Scattering from Condensed
Matter}, (Oxford University Press, 1984).


\bibitem{Kern-PRB-1985}
S. Kern {\it et al}, Phys. Rev. B {\bf 32}, 3051 (1985).

\bibitem{Kern-JAP-1990}
S. Kern {\it et al}, J. Appl. Phys. {\bf 67}, 4830 (1990).

\bibitem{footnote2}
The $B_{q}^{k}$ and $\zeta$ coefficients used here are defined in
W.T. Carnall {\it et al}, J. Chem. Phys. {\bf 90}, 3443 (1989).

\bibitem{Kaufman-JRNBS-1967}
V. Kaufman and J. Sugar, J. Res. NBS {\bf 71A}, 583 (1967).

\bibitem{PJB-WBL}
P.J. Brown, in {\it International Tables for X-ray
Crystallography}, edited by A.J.C. Wilson (Kluwer Academic,
Dordrecht, 1992), Vol. C; W.B. Lewis, in {\it Magnetic Resonance
and Related Phenomena}, edited by I. Ursu (Proc. XVIth Congress
AMPERE, Bucharest, 1970), p. 717.

\bibitem{footnote3}
These are somewhat smaller than the corresponding values 192 and
118\,mb\,sr$^{-1}$Pr$^{-1}$ obtained when $J$-mixing with the
$J=\frac{7}{2}$ multiplet is neglected.

\bibitem{footnote4}
A further $\sim$16\,mb\,sr$^{-1}$Pr$^{-1}$ associated with elastic
(Bragg) scattering from the ordered moment of
$\sim$0.50\,$\mu_{\rm B}$ at 10\,K is not contained in the data in
Fig.\ 1(d).

\bibitem{Sasaki}
K. Sasaki and Y. Obata, J. Phys. Soc. Jpn. {\bf 28}, 1157 (1970).

\bibitem{Sugar-PRL-1965}
J. Sugar, Phys. Rev. Lett. {\bf 14}, 731 (1965).





\end{references}
\end{document}